\documentclass[aps,pra,showpacs,notitlepage,superscriptaddress,twocolumn]{revtex4-1}

\usepackage[utf8x]{inputenc}
\usepackage{amsmath}
\usepackage{amssymb}
\usepackage{amsfonts}
\usepackage{graphics}
\usepackage[pdftex]{graphicx}
\usepackage[normal]{subfigure}
\usepackage{color}
\usepackage{fullpage}

\begin{document}

\title{Quantum walks and non-Abelian discrete gauge theory}

\author{Pablo Arnault}
\email{pablo.arnault@upmc.fr}
\affiliation{LERMA, Observatoire de Paris, PSL Research University, CNRS, Sorbonne Universit\'es, UPMC Univ. Paris 6, UMR 8112, F-75014, Paris, France}

\author{Giuseppe Di Molfetta}
\email{giuseppe.di.molfetta@uv.es}
\affiliation{Departamento de F{\'i}sica Te{\'o}rica and IFIC, Universidad de Valencia-CSIC, 46100-Burjassot, Valencia, Spain}

\author{Marc Brachet}
\affiliation{CNRS, Laboratoire de Physique Statistique, {\'E}cole Normale Sup\'erieure, 75231 Paris Cedex 05}

\author{Fabrice Debbasch}
\affiliation{LERMA, Observatoire de Paris, PSL Research University, CNRS, Sorbonne Universit\'es, UPMC Univ. Paris 6, UMR 8112, F-75014, Paris, France}

\date{\today}

\begin{abstract}


A new family of discrete-time quantum walks (DTQWs) on the line with an exact discrete $U(N)$ gauge invariance is introduced. It is shown that the continuous limit of these DTQWs, when it exists, coincides with the dynamics of a Dirac fermion coupled to usual $U(N)$ gauge fields in $2D$ spacetime. A discrete generalization of the usual $U(N)$ curvature is also constructed. An alternate interpretation of these results in terms of superimposed $U(1)$ Maxwell fields and  
$SU(N)$ gauge fields is discussed in the Appendix. Numerical simulations are also presented, which explore the convergence of the DTQWs towards their continuous limit and which also compare the DTQWs with classical ({\sl i.e.} non-quantum) motions in classical $SU(2)$ fields. The results presented in this article constitute a first step towards quantum simulations of generic Yang-Mills gauge theories through DTQWs.


\end{abstract}

\maketitle

\section{Introduction}

Discrete-time quantum walks (DTQWs) are unitary quantum automata and can be viewed as formal generalizations of classical random walks. They were first considered in a systematic way by Meyer \cite{Meyer96a}, following the seminal work of Feynman \cite{FeynHibbs65a} and Aharonov \cite{ADZ93a}. DTQWs have been realized experimentally  with a wide range of physical objects and setups \cite{Schmitz09a, Zahring10a, Schreiber10a, Karski09a, Sansoni11a, Sanders03a, Perets08a}, and are studied in a large variety of contexts, ranging from fundamental quantum physics \cite{Perets08a, var96a} to quantum algorithmics \cite{Amb07a, MNRS07a}, solid-state physics \cite{Aslangul05a, Bose03a, Burg06a, Bose07a} and biophysics \cite{Collini10a, Engel07a}.

It has been shown recently that the continuous limit of several DTQWs coincides with the dynamics of Dirac fermions coupled to electromagnetic \cite{di2012discrete, ADmag16, ADelectromag16} and relativistic gravitational fields \cite{DMD13b,DMD14,Arrighi_curved_1D_15,SFGP15a}. Though these fields are naturally gauge fields, they are not generic Yang-Mills gauge fields. Indeed, electromagnetism is based on the Abelian 
gauge group $U(1)$, while relativistic gravitational fields are not Yang-Mills gauge fields, since they are represented by a metric, and not by a connection. The aim of this article is to exhibit and study DTQWs whose continuous limit coincides with the dynamics of a fermion coupled to Yang-Mills $U(N)$ gauge fields.


To make things definite and as simple as possible, we focus on $1D$ DTQWs. The minimal $1D$ DTQWs have a two-dimensional coin space. Their wave functions thus have two components, one propagating towards the left and one towards the right. To take into account the internal degrees of freedom associated to $U(N)$ gauge invariance, we consider $1D$ DTQWs with coin space of dimension $2N$ {\sl i.e.}  $2N$-component wave functions. Half of the wave-function components propagates towards the left, and the other half towards the right. The so-called mixing operator advancing the walk in time is represented by a $2N \times 2N$ time- and space-dependent unitary matrix. 

We introduce new $1D$ DTQWs with $2N$ components which admit an exact discrete $U(N)$ gauge invariance and build for the DTQWs a discrete equivalent $\mathcal F$ of the usual Yang-Mills curvature $F$.
We then prove that the limit of these DTQWs, when it exists,  coincides with the dynamics of Dirac fermions coupled to $U(N)$ gauge fields and that the discrete curvature $\mathcal F$ tends towards $F$ in the continuous limit.

These formal computations are complemented by numerical simulations. These address the convergence of the DTQWs towards their continuous limit and the correspondence with classical ({\sl i.e.} non-quantum) trajectories in Yang-Mills fields \cite{Boozer11}. The article concludes by a brief summary and a discussion of the main results. Finally, the Appendix elaborates on the fact that a $U(N)$ gauge field can be viewed as the superposition of a $U(1)$ Maxwell field and an $SU(N)$ gauge field, and reinterprets our results in that alternate context.

\section{The DTQWs and their gauge invariance}

\subsection{The DTQWs}

We consider DTQWs defined over discrete time and infinite discrete one-dimensional space. Instants are labelled by $j \in \mathbb{N}$  and space points by  $p \in \mathbb{Z}$. The coin space of the DTQWs has dimension $2N$. Given a certain orthonormal basis in this space, the wave functions $\Psi$ of the walks are represented by $2N$ components and we group these components into two $N$-component sets $\psi^-$ and $\psi^+$, which represent those parts of $\Psi$ which propagate respectively to the left and to the right. 
The evolution equations read
\begin{equation}
\begin{bmatrix} \psi^{-}_{j+1, p}\\ \psi^{+}_{j+1, p } \end{bmatrix} \  = 
\mathbf{B}\left(\theta, P_{j, p} , Q_{j, p} \right)
 \begin{bmatrix} \psi^{-}_{j, p +1  } \\ \psi^{+}_{j, p -1 } \end{bmatrix} ,
\label{eq:defwalkdiscr}
\end{equation}
with
\begin{align}
\mathbf{B}(\theta, P,Q) &=  \left( 
\mathbf{C}(\theta)
\otimes \mathbf{1}_N \right) \times 
\begin{bmatrix} P &  0  \\  0  &   Q
 \end{bmatrix} \\
&\equiv \begin{bmatrix}
 (\cos \theta) \, P & (i \sin \theta) \,  Q \\
 (i \sin \theta )  \, P & (\cos \theta) \,  Q
 \end{bmatrix} , \nonumber
\label{eq:defB}
\end{align}
where 
$\otimes$ is the so-called Kronecker (or tensorial) product for matrices, and $P$, $Q$ are elements of $U(N)$. 
These walks are unitary {\sl i.e.} $\Pi_j = \sum_{p}|\Psi_{j,p}|^2$ is independent of $j$.

In the continuous limit, the parameter $\theta$ will code for the mass of the fermion and the matrices $P$ and $Q$ will code for the potential of the $U(N)$ gauge field to which the fermion is coupled. 

\subsection{Discrete $U(N)$ gauge invariance and discrete curvature}

The DTQWs defined by (\ref{eq:defwalkdiscr}) admit a discrete local $U(N)$ gauge invariance. Indeed, consider the local gauge transformation $\Psi_{j,p}$ = $( \mathbf{1}_2 \otimes G^{-1}_{j,p})$ $\Psi'_{j,p}$, where $G_{j,p}$ is some matrix of $U(N)$. Equations (\ref{eq:defwalkdiscr}) are kept invariant under this transformation, that is
\begin{equation}
\begin{bmatrix} {\psi'}^{-}_{j+1, p} \\ {\psi'}^{+}_{j+1, p} \end{bmatrix} \  = 
\mathbf{B}\left( \theta, P'_{j, p} ,  Q'_{j, p} \right)
\begin{bmatrix} {\psi'}^{-}_{j, p+1  } \\ {\psi'}^{+}_{j, p-1 } \end{bmatrix},
\label{eq:TransWalk}
\end{equation}
provided that we set
\begin{align}
P'_{j,p}  &= G_{j+1,p} \, P_{j,p} \, G^{-1}_{j, p+1} \nonumber \\
Q'_{j,p}  &= G_{j+1,p} \, Q_{j,p} \, G^{-1}_{j, p-1} \ .
\label{eq:gaugeRel}
\end{align}

The above gauge invariance suggests that $R = (P, Q)$
is the discrete equivalent of the usual continuous $U(N)$ gauge potentials. This will be confirmed in Section \ref{sec:continuous_limit}, where the continuous limit of (\ref{eq:defwalkdiscr}) will be derived. We now wish to build out of $R$
an object $\mathcal F$ defined on the spacetime lattice, which generalizes for DTQWs the usual curvature (field-strength) tensor \cite{Quigg} $F$ of standard gauge fields. This will be done by searching for an object whose transformation law under a change of gauge ressembles the transformation law of $F$.
Let 
\begin{align}
U_{j,p} (R) &= Q^{\dag}_{j,p} \, P_{j,p} \nonumber \\
V_{j,p} (R) &= Q_{j,p}  \, P_{j-1, p-1} \ ,
\end{align}
whose transformation under a change of gauge reads,
\begin{align}
U'_{j,p} (R') &= G_{j, p-1} \, U_{j,p} (R) \, G_{j, p+1}^{\dag} \nonumber \\
V'_{j,p} (R') &= G_{j+1, p} \, V_{j,p} (R) \, G_{j-1, p}^{\dag} \ , 
\end{align}
involving shifts of $G_{j,p}$ only in the spatial ({\sl resp.} temporal) dimension for $U_{j,p}(R)$ ({\sl resp.} $V_{j,p}(R)$), while these shifts were mixed in the transformation laws of Eq. (\ref{eq:gaugeRel}). From these equations, we can write transformation laws involving the $2 \times 2 = 4$ discrete-spacetime neighbours of $G_{j,p}$, 
\begin{align}
U'_{j+1, p} (R') &= G_{j+1, p-1} \, U_{j+1, p} (R) \, G_{j+1, p+1}^{\dag} \nonumber \\
U'_{j-1, p} (R') &= G_{j-1, p-1} \, U_{j-1, p} (R) \, G_{j-1, p+1}^{\dag} \\
V'_{j, p-1} (R') &= G_{j+1, p-1} \, V_{j, p-1} (R) \, G_{j-1, p-1}^{\dag} \nonumber \\
V'_{j, p+1} (R') &= G_{j+1, p+1} \, V_{j, p+1} (R) \, G_{j-1, p+1}^{\dag} \ , \nonumber
\end{align}
from which we can build
\begin{align}
{\mathcal F}_{j,p} (R)
= U_{j-1, p}^{\dag}(R) \, V_{j, p-1}^{\dag}(R)\,  U_{j+1, p}(R) \, V_{j, p+1}(R) \, ,
\label{eq:THE_matrix}
\end{align} 
whose transformation law reads
\begin{equation} \label{eq:gauge_transfo_rule}
{\mathcal F}'_{j,p} (R') = 
 G_{j-1, p+1} \, {\mathcal F}_{j,p}(R) \, G^{-1}_{j-1, p+1} \ .
\end{equation}
As will become apparent in the next section, ${\mathcal F}$ is a discrete equivalent to the curvature (field-strength) tensor of continuous gauge fields.

\section{Continuous limit} \label{sec:continuous_limit} \label{sec:continuous_limit}

We now show that it is possible to choose $\theta$, $P$ and $Q$ in such a way that (\ref{eq:defwalkdiscr}) admits a continuous limit identical to the Dirac equation for a fermion coupled to an arbitrary $U(N)$ gauge field.

In order to compute the continuous limit of equation (\ref{eq:defwalkdiscr}), we first introduce a dimensionless time and space step $\epsilon$, and consider that $\Psi_{j,p}$, $P_{j,p}$, $Q_{j,p}$ 
are the values $\Psi(t_j,x_p)$,  $P(t_j,x_p)$ and $Q(t_j,x_p)$
taken at spacetime point $(t_j = j \epsilon, x_p = p \epsilon)$ by a $2N$-component wave function $\Psi$ and two time- and spacetime-dependent matrices $P$ and $Q$ in $U(N)$.
 We then assume that $\Psi$, $P$ and $Q$ 
 are at least twice differentiable with respect to both space and time variables and let $\epsilon$ tend to zero.

As $\epsilon$ tends to zero, the wave functions on the left-hand side and on the right-hand side of (\ref{eq:defwalkdiscr}) both tend towards 
$\Psi(t_j, x_p)$. Thus, the continuous limit of (\ref{eq:defwalkdiscr}) can only exist if, in that limit, $B(\theta, P, Q)$ tends to unity at all points in spacetime. 
This is achieved by choosing an angle $\theta$ which tends to zero with $\epsilon$ and two matrices $P$ and $Q$ which tend to unity as $\epsilon$ goes to zero. 
We retain $\theta = - \epsilon m$, where $m$ is a positive constant (as opposed to a function of $t$ and $x$) which will play the role of a mass in the continuous limit.
As for the matrices $P$ and $Q$, we remark that $U(N)$ is a compact and connected Lie group. Thus, the exponential map generates the whole group \cite{Cornwell} {\sl i.e.} all elements $M\in U(N)$ can be written as 
\begin{equation}
M = \exp \left( 
i \sum_k X^k_M \tau_k
\right),
\end{equation}
where the $\tau_k$'s are $N^2$ generators of $U(N)$ and the $X^k_M$'s can serve as coordinates for $M$.

To ensure that both functions $P(t, x)$ and $Q(t, x)$ tend to unity as $\epsilon$ goes to zero, we choose $X^k_P (t, x) = \epsilon b^k_P(t, x)$  and $X^k_Q (t, x) = \epsilon b^k_Q(t, x)$,
where  $b^k_{P/Q}(t, x)$ are two real functions independent of $\epsilon$.

Taylor expanding (\ref{eq:defwalkdiscr}) at first order in $\epsilon$ and letting $\epsilon$ tend to zero then delivers
{\small
\begin{align} \label{eq:dirac2}
 (\partial_0 - i b^k_0 \tau_k) \psi^- - (\partial_1  - i b^k_1 \tau_k) \psi^- &= -i  m \psi^+   \\
 (\partial_0 - i b^k_0 \tau_k)  \psi^+ + (\partial_1 - i b^k_1 \tau_k) \psi^+ &= - i m \psi^- , \nonumber
\end{align}}
where $\partial_0= \partial_t$, $\partial_1= \partial_x$,
\begin{align}
b_0 &= (b_Q + b_P)/2 \nonumber \\
b_1 &= (b_Q - b_P)/2 \ ,
\end{align}
and summation over repeated index $k$ is implied.

Equations (\ref{eq:dirac2}) can be recast as
\begin{equation}
 \left[i \gamma^\mu D_\mu - m \right]\Psi= 0 \ ,
\label{eq:yangmillseq}
\end{equation}
where index $\mu$ is summed over from $0$ to $1$, with the gamma matrices $\gamma^0 = \sigma_1 \otimes \mathbf{1}_N$,  $\gamma^1 = i\sigma_2 \otimes \mathbf{1}_N$, and the covariant derivative $D_\mu = \partial_\mu - i b^k_\mu \tau_k$. Equation (\ref{eq:yangmillseq}) is the flat-spacetime Dirac equation, with convention $[\eta^{\mu\nu}]= \text{diag}(+,-)$, for a spin-1/2 fermion of mass $m$ coupled to a non-Abelian $U(N)$ potential $b^k_\mu \tau_k$ (with coupling constant $g = -1$)
\cite{Quigg} belonging to the Lie algebra of $U(N)$.
Note that the $b^k_\mu$'s are real-valued space- and time-dependent fields.

Taylor expanding Definition (\ref{eq:THE_matrix}) for $\mathcal F$ delivers
\begin{equation}
{\mathcal F} (t, x) = \mathbf{1}_N + 4 \epsilon^2 F_{10} (t, x) + \mathcal{O}(\epsilon^3) \ ,
\end{equation}
where $F_{10}$ is the only non-vanishing component of the antisymmetric curvature (field-strength) tensor $F_{\mu \nu}$ of the connection $B_\mu = b^k_\mu \tau_k$, defined by
\begin{equation} \label{eq:Yang_Mills_tensor}
F_{\mu\nu} = \partial_{\mu}B_{\nu} - \partial_{\nu} B_{\mu} -i[B_{\mu},B_{\nu}] \ ,
\end{equation}
with $[B_{\mu},B_{\nu}] = B_{\mu}B_{\nu} - B_{\nu}B_{\mu}$.
Note that the transformation law for $F_{\mu \nu}$ under a change of gauge reads
\begin{equation} \label{eq:continuous_gauge_transfo}
F'_{\mu\nu} = G F_{\mu\nu} G^{-1} \ ,
\end{equation}
which closely parallels (\ref{eq:gauge_transfo_rule}).

\section{Numerical simulations of $U(2)$-invariant DTQWs} 


\subsection{Simulated walk} \label{sec:SU2}

As shown in the Appendix to this article, $U(N)$ factorizes into the product of $U(1)$ and $SU(N)$. In physical terms, this means that a $U(N)$ gauge field can be seen as the superposition of a $U(1)$ Maxwell field and an 
$SU(N)$ gauge field. The effects of Maxwell fields on DTQWs have already been presented in several publications \cite{di2012discrete, ADmag16, DMD14}. We want to focus on the effects of non-Abelian Yang-Mills fields and thus choose to simulate situations where the Maxwell field identically vanishes so that the $U(N)$ gauge field is then actually an $SU(N)$ gauge field. We also choose the simplest option $N = 2$. The group $SU(2)$ is compact and connected, and can thus be fully generated by the exponential map, from three generators ${\bar \tau}_k$, $k = 1, 2, 3$, belonging to its Lie algebra. We retain (see Appendix) ${\bar \tau}_k=\sigma_k/2$ where the $\sigma_k$'s are the three Pauli matrices, and choose 

\begin{align} \label{eq:potentials}
{\bf{\bar X}}_0 &= ({\bf{\bar X}}_Q + {\bf{\bar X}}_P)/2 = (0,0,0) \nonumber \\
{\bf{\bar X}}_1 &= ({\bf{\bar X}}_Q - {\bf{\bar X}}_P)/2 = (\epsilon E_{\text{YM}}t,0,0) \ .
\end{align}
The bar is used to distinguish the notations used for $SU(2)$ from those used for $U(N)$, including $U(2)$, in Section \ref{sec:continuous_limit}. The boldface notation is used as a reminder that the Lie algebra of $SU(2)$ is of dimension $3$. 
The continuous limit can be recovered by letting $\epsilon$ tend to zero (see Section \ref{sec:continuous_limit}). In such a continuous limit, this potential, (\ref{eq:potentials}), generates a uniform and constant $SU(2)$ `electric' field $E_{\text{YM}}$ in the ${\bar \tau}_1$ direction of the $SU(2)$ Lie algebra.

\subsection{Convergence towards the continuous limit}
To study numerically the convergence of a DTQW towards a solution of the Dirac equation, we chose an initial wave function and compare, for some given time $j$, its evolution  $\psi^{u}_{QW}(t_j=\epsilon j,\cdot), u \in \lbrace +, - \rbrace$, through the DTQW to the evolution $\psi^{u}_{D}(t_j,\cdot)$ of the same initial condition through the Dirac equation.
The comparison is carried out through the following mean relative difference,
\begin{equation}
 \small{\delta \psi^{u}_{j} = \sqrt{ \frac{\langle |\psi^u_D(t_j,\cdot)-\psi^u_{QW}(t_j,\cdot)|^2\rangle}{\langle |\psi^u_D(t_j,\cdot)|^2\rangle}}} \ ,
\end{equation}
where
 \begin{equation} \label{eq:L2_norm}
\langle f(t_j,\cdot) \rangle =  \sum_{p = -p_{\mathrm{max}}(\epsilon)}^{p_{\mathrm{max}}(\epsilon)}  f(t_j,x_p) \ \epsilon \ ,
\end{equation}
The numerical simulations are carried out over the space interval $[-x_{\text{max}},x_{\text{max}}]$ with $x_{\text{max}}=200$, and $p_{\text{max}}(\epsilon) \equiv x_{\text{max}}/\epsilon$. The maximal time over which we carry out the simulations, $t_{\text{max}}=350$, is short enough so that the walker never reaches the spatial boundaries.

Note that $\delta \psi^{u}_{j}$ does not measure the difference between quantum states, for which phase differences are unimportant, but rather the difference between the functions $\psi^u_D$ and $\psi^u_{QW}$. This is appropriate here because we want to test the convergence of a discrete scheme towards its formal continuous limit, and this convergence should be verified on both modulus and phase {\sl i.e.} on the whole complex function, and not only on the state it represents.

Since there is only a time dependence and no space dependence in the potentials (\ref{eq:potentials}), we can use as numerical solver for the Dirac equation standard pseudo-spectral methods \cite{gottlieb1977numerical, guo2003spectral}, with resolution $2 \pi/\epsilon$ in $2 \pi$-periodic boundary conditions. Time marching is performed using a second-order Runge-Kutta scheme. The original DTQW can also be simulated in spectral space using the standard translation operator in Fourier space. 
\begin{figure}[h!]
\includegraphics[width=1.07\columnwidth]{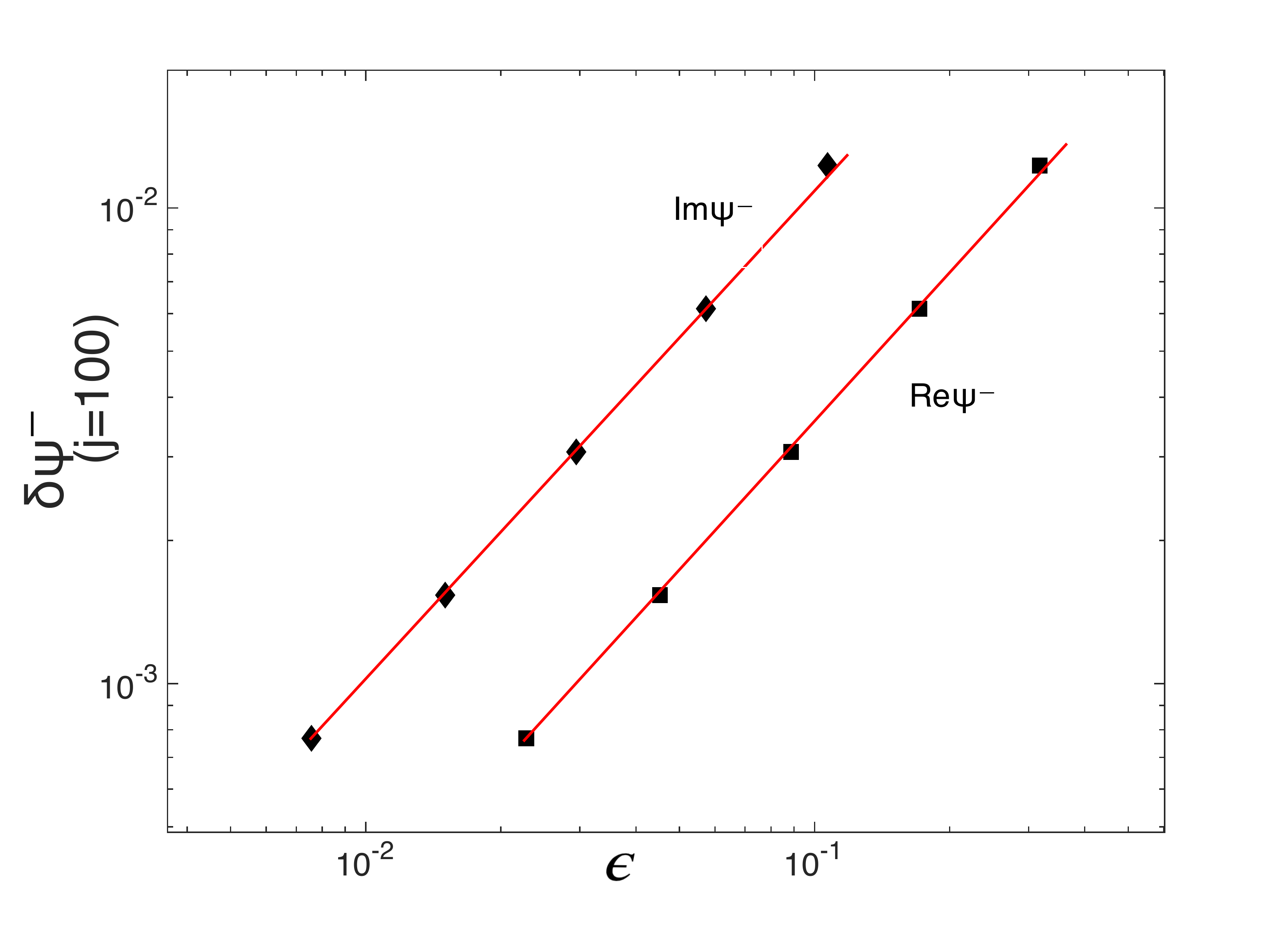}
\caption{(Color online) Relative differences $\delta f_j$ for $f= \text{Im} \, \psi^-$ and $f= \text{Re} \, \psi^-$ as functions of $\epsilon$, at time $j = 100$, with $m$ = $0.1$ and $E_{\text{YM}}=0.08$. The initial condition is given by Eq. (\ref{eq:gauss}) with $\sigma = 0.5$. }
\label{fig:RelativeDistance}
\end{figure}

Figure \ref{fig:RelativeDistance} shows that the mean relative differences $\delta f_j$ for $f= \text{Im} \, \psi^- $ and $\text{Re} \, \psi^-$, scale as $\epsilon$ as expected: indeed, this scaling coincides with the theoretical expectation since, 
for a single time step, the discrepancy is theoretically of order $\epsilon^2$. Thus, after a fixed time $t=\mathcal{O}(\epsilon^{-1})$, the discrepancy is of order $\epsilon^{-1} \epsilon^2$= $\epsilon$. These results also confirm that the DTQW (\ref{eq:defwalkdiscr}) with choice (\ref{eq:potentials}) can be used to simulate massive Dirac dynamics in a constant and uniform non-Abelian `electric' field $E_{\text{YM}}$.

\subsection{Comparison with classical trajectories}

Given a wave equation, it is well known \cite{LandauMQ} that the center of mass of a wave-packet solution follows classical trajectories. 
In the continuous-limit case described above in section \ref{sec:SU2}, the corresponding classical equations have been explicitly derived in \cite{Boozer11}.
We now want to investigate whether the original DTQW also reproduces classical motions of the center of mass of wave packets.

We consider $k_0$-centered Gaussian wave packets of positive-energy eigenvector $u_+(k)$ of the two-component ({\sl i.e.} without $SU(2)$ internal degree of freedom) free Dirac Hamiltonian,
tensorised with an equally weighted initial $SU(2)$ state:
\begin{equation}
\Psi(x) = \int dk \ e^{-\frac{(k-k_0)^2}{2\sigma^2} + i x k} (u_+(k) \otimes (1,1)^{\top}/\sqrt{2}) \ ,
\label{eq:gauss}
\end{equation} 
where superscript $\top$ denotes the transposition, and	
{\small
\begin{align}
u_+(k) &= \nonumber \\
&\hspace{-0.7cm} \left[
 \frac{\sqrt{k^2+m^2}-k}{ \sqrt{\left(\sqrt{k^2+m^2}-k\right)^2+m^2}},
\frac{1}{\sqrt{\frac{\left(\sqrt{k^2+m^2}-k\right)^2}{m^2}+1}}
\right]^{\top}.
\end{align}}

\begin{figure}[h]
\includegraphics[width=1.08\columnwidth]{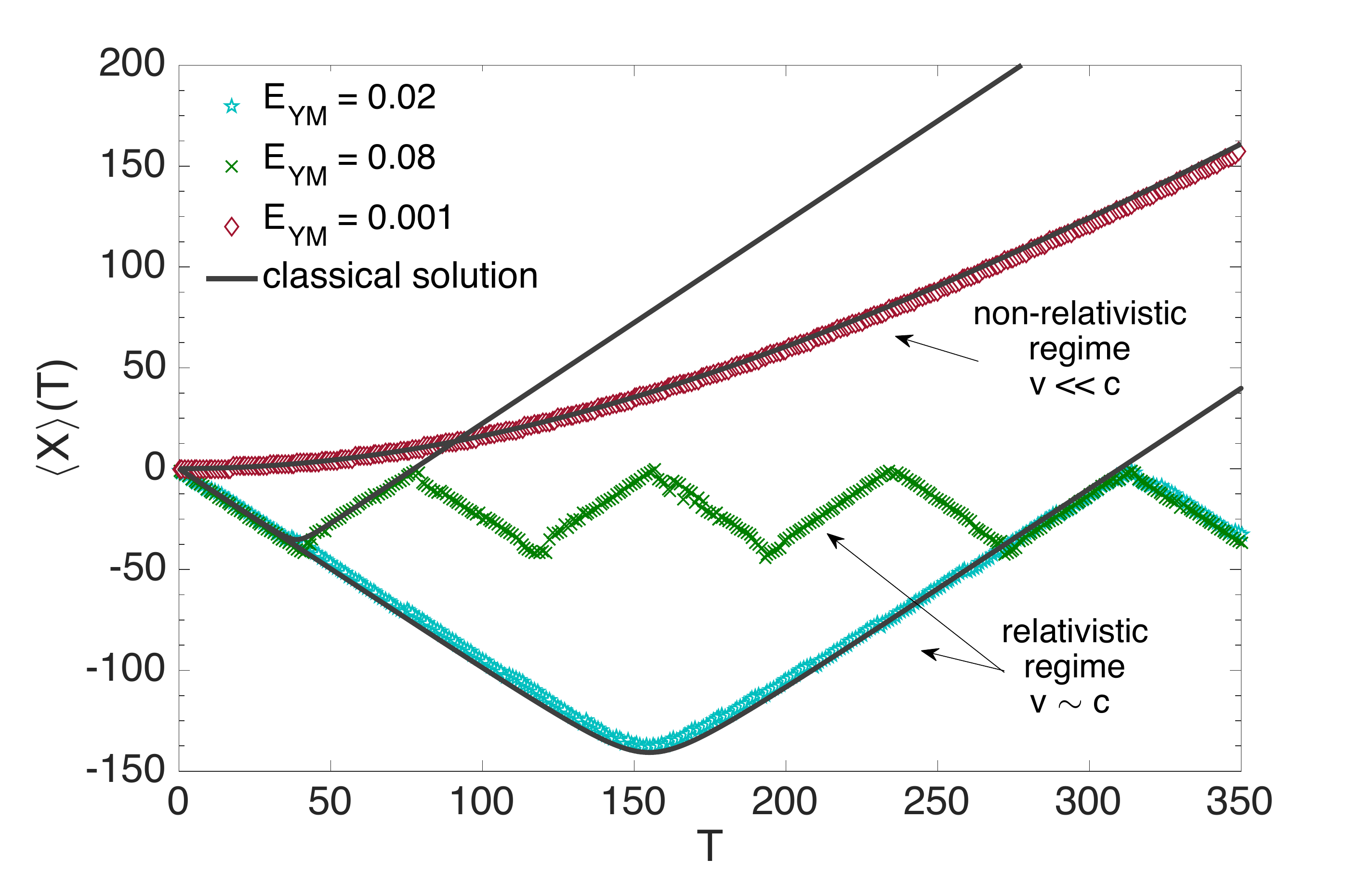}
\caption{(Color online) 
Time evolution of the DTQW mean trajectory $\bar{x}(t)$ for a non-Abelian coupling constant $g=1$ and different values of $E_{\text{YM}}$, versus classical trajectories (black solid line). Short-time agreement between quantum and classical dynamics is shown in the ultra-relativistic range $k_0^2 = 1$ (green and blue) and the non-relativistic range $k_0^2 = 0$ (red). The initial condition is given by Eq. (\ref{eq:gauss}) with $\sigma = 0.5$ (green and blue), $\sigma = 1$ (red) and $m$ = $0.1$. }
\label{fig:traj}
\end{figure}

Figure \ref{fig:traj} demonstrates the short-time agreement between solutions of classical particle trajectory equations (see Ref. \cite{Boozer11}) and the centers of wave packets  $\bar{x}(t)$ obtained from DTQW solutions, both in the non-relativistic, $k_0^2=0$, $m=0.1$, and in the relativistic case, $k_0^2=1$, $m=0.1$. When the agreement is lost, oscillatory trajectory for  $\bar{x}(t)$ are produced by the DTQW. Note that similar long-time oscillations are also found in the simple context of DTQWs corresponding to Dirac fermions coupled to electromagnetic fields \cite{witthaut2010quantum}.

\section{Conclusion}
We have introduced new DTQWs on the line which exhibit an exact discrete $U(N)$ gauge invariance and whose continuous limit coincides, when it exists, with the dynamics of Dirac fermions coupled to $U(N)$ gauge fields. We have also built a discrete generalization of the curvature tensor of the gauge fields. We have finally complemented these analytical results by numerical simulations which explore the convergence of the DTQWs towards their continuous limit and compare the DTQWs with the dynamics of non-quantum particles in classical gauge fields. The interpretation of our results in terms of Maxwell fields superimposed to $SU(N)$ gauge fields is presented in the Appendix. The results presented in this article constitute a first step towards quantum simulations of generic Yang-Mills gauge theories through DTQWs. Until now, only DTQWs with two-component wave functions have been realized experimentally \cite{Genske_Meschede}. But experimental procedures allowing the implementation of DTQWs with wave functions having more than two components have been proposed in \cite{NRS03a,GRFK15}. In these procedures, the DTQWs are implemented with single photons or classical light, for example in optical cavities.

Let us now mention of few avenues open to future studies. The DTQWs presented in this article should first be extended to $(1 + 2)$, and then to $(1 + 3)$ spacetime dimensions. Note that DTQWs modeling Dirac fermions coupled to $U(1)$ gauge fields have already been proposed in $(1 + 1)$ and $(1 + 2)$ dimensions \cite{di2012discrete, ADmag16, ADelectromag16}. Another possible extension would be the construction of DTQWs which are coupled, not only to $U(N)$ gauge fields, but also to gravity. Until now, this has only been done for $N = 1$ and in $(1 + 1)$ spacetime dimensions \cite{DMD14}. Also, performing full quantum simulations of Yang-Mills gauge theories will require complementing the fermionic DTQW dynamics by dynamical equations for the discrete gauge fields {\sl i.e.} for matrices $P, Q \in U(N)$ which define the DTQW. The dynamical equations for the gauge field should be a set of finite difference equations relating the discrete curvature (field-strength) tensor $\mathcal F$ introduced in the present article to a discrete gauge-invariant fermionic current associated to the DTQW. This current has already been presented in \cite {ADelectromag16} for $N = 1$ in $(1 + 2)$ spacetime dimensions, and the corresponding discrete Maxwell equations have also been written down. The procedure should now be extended to generic non commutative discrete gauge fields. Finally, incorporating Yang-Mills fields to DTQWs defined on arbitrary graphs is certainly worth working on, if only for applications to quantum information.


\section{Acknowledgements}
This work has been partially supported by the Spanish Ministerio de Educaci{\'o}n e Innovaci{\'o}n, MICIN-FEDER project  FPA2014-54459-P, SEV-2014-0398 and `Generalitat Valenciana' grant GVPROMETEOII2014-087. 



\appendix

\section{Alternative physical interpretation}

The Lie group $U(N)$ is the group of $N \times N$ unitary matrices {\sl i.e.} $N \times N$ matrices whose determinant modulus equals unity. In particular, elements of $U(1)$ are complex numbers of unit modulus {\sl i.e.} complex numbers of the form $\exp( i \beta)$, where $\beta$ is an arbitrary real number. The group $U(N)$ is Abelian for $N = 1$ and non Abelian for $N > 1$. Consider now an arbitrary element $M$ of $U(N)$, 
its determinant $\mbox{det} M = \exp(i \alpha)$, $\alpha \in \big]-\pi, + \pi \big]$, and we define the matrix ${\bar M} = M/\delta$ where $\delta^N = \mbox{det} M$. The matrix ${\bar M}$ has unit determinant and is thus an element of the special unitary group $SU(N)$. The group $U(N)$ can therefore be factorized into the direct product of $U(1)$ and $SU(N)$. This factorization is not unique because $\delta$ is not uniquely defined by Equation $\delta^N = \mbox{det} M$. 
Indeed, this equation has the $N$ distinct solutions $\delta_k = \exp \left[ i( \alpha + 2k \pi)/N\right]$, $k = 0, ..., N-1$, and each solution defines a different factorization. Note also that imposing a factorization which depends continuously on $M$ is only possible if one makes a cut along the negative real axis in the complex plane of $\mbox{det} M$ {\sl i.e.} if one does not define the factorization for matrices $M$ whose determinant  corresponds to the value $\alpha = \pi$ (and is thus equal to $-1$).

To make all computations definite, we now choose $k = 0$ in the above definition of $\delta_k$. This defines unambiguously a factorization of $U(N)$ into the direct product of $U(1)$ and $SU(N)$. This factorization is not continuous for matrices $M$ with $\mbox{det} M = -1$, but that should not be a practical problem when one is working on a spacetime lattice. In the continuous limit, all $U(N)$ matrices considered in this article tend to unity. Their determinant is thus close to unity and the retained factorization is thus defined and continuous for all these matrices.

In physical term, the existence of the factorization means that a $U(N)$ gauge field can be interpreted as the superposition of a $U(1)$ Maxwell field and an $SU(N)$ gauge field. Now, $SU(N)$ is itself a compact and connected Lie group, so the whole group is generated from the identity by the exponential map. The above factorization can thus be used to write all matrices $M \in U(N)$ as
\begin{equation}
M = \delta_M {\bar M} = \exp(i Y_M) \exp (i \sum_k {\bar X}^k_M {\bar \tau}_k) \ ,
\end{equation}
where the ${\bar \tau}_k$'s are the $N^2 - 1$ generators of $SU(N)$. This point of view is adopted in Section \ref{sec:SU2}. Note that the factorization of $U(N)$ also shows that the DTQWs presented in this article coincide, for $N = 1$, with the DTQWs already proposed to simulate Dirac fermions coupled to arbitrary electric fields \cite{di2012discrete}.

The discrete curvature $\mathcal F$ also factorizes into a curvature for the Maxwell field and a curvature for the $SU(N)$ gauge field. One finds indeed that
\begin{equation}\label{eq:discrete_strength_tensor}
{\mathcal F}_{j,p}(R) = 
{\mathcal F}_{j,p}(\delta_R) 
\, {\mathcal F}_{j,p}({\bar R})
\end{equation}
where  $\delta_R = (e^{i Y_P},e^{i Y_Q})$ and ${\bar R} = ({\bar P}, {\bar Q})$. The $SU(N)$ curvature ${\mathcal F}_{j,p}({\bar R})$ is given by (\ref{eq:THE_matrix}) and the $U(1)$ Abelian curvature reads
\begin{equation}
{\mathcal F}_{j,p}(\delta_R) = \exp \left[2 i (\mathcal{I} f_{10})_{j,p} \right] \ ,
\end{equation}
where
\begin{equation}\label{eq:abelian_discrete_strength_tensor}
(f_{10})_{j,p} = (d_{1} Y_0)_{j,p} - (d_{0} Y_1)_{j,p} \ ,
\end{equation}
with
\begin{align}
Y_0 &= (Y_Q + Y_P)/2 \nonumber \\
Y_1 &= (Y_Q - Y_P)/2,
\end{align}
\begin{align}
d_0 = (L_0 - \Sigma_1)
, \ \ \ \ \  d_1 = \Delta_1 ,
\end{align}
and
\begin{align}
(L_0 K)_{j,p} &= K_{j+1,p} \nonumber \\
(\Sigma_1 K)_{j,p} &= (K_{j,p+1} + K_{j,p+1})/2 \\
(\Delta_1 K)_{j,p} &= (K_{j,p+1} - K_{j,p+1})/2 \ , \nonumber
\end{align}
where $K$ is an arbitrary quantity which depends on $j$ and $p$.
Operator $\mathcal I$ is defined in terms of $L_0$ and $L_1$ by
\begin{equation}
\mathcal{I} = 1 + L_0^{-1} L_1^{-1} \ .
\end{equation}
This form of ${\mathcal F}_{j,p}(\delta_R)$ is interesting because $(f_{10})_{j,p}$ and operators $d_0$, $d_1$ have already been introduced in \cite{ADelectromag16} in the context of DTQWs exhibiting a $U(1)$ gauge invariance.

Choosing $Y_\mu = \epsilon A_\mu$, ${\bar X}^k_{P/Q} = \epsilon {\bar b}^k_{P/Q}$ and Taylor expanding (\ref{eq:discrete_strength_tensor}) at second order in $\epsilon$ delivers
\begin{equation}
{\mathcal F}_{j,p}(R) = \mathbf{1}_{N} + 4 \epsilon^2 (f_{10}\mathbf{1}_{N} + {\bar F}_{10}) + \mathcal{O}(\epsilon^3) \ ,
\end{equation}
where ${ \bar F}_{10}$ is given by (\ref{eq:Yang_Mills_tensor}) after substitution $B \rightarrow \bar{B}$, and $f_{10} = \partial_1 A_0 - \partial_0 A_1$ is the ($10$)-component of the usual Abelian 
curvature tensor.

\end{document}